# Analog and digital circuit design in 65 nm CMOS: end of the road?

Georges Gielen, K.U. Leuven, Belgium
Wim Dehaene, K.U.Leuven, Belgium
Phillip Christie, Philips, The Netherlands
Dieter Draxelmayr, Infineon, Austria
Edmond Janssens, ST Microelectronics, Belgium
Karen Maex, IMEC, Belgium
Ted Vucurevich, Cadence, USA

## Abstract

*This special session adresses the problems that designers face when implementing analog and digital circuits in nanometer technologies. An introductory embedded tutorial will give an overview of the design problems at hand : the leakage power and process variability and their implications for digital circuits and memories, and the reducing supply voltages, the design productivity and signal integrity problems for embedded analog block s. Next, a panel of experts from both industrial semiconductor houses and design companies, EDA vendors and research institutes will present and discuss with the audience their opinions on whether the design road ends at marker "65nm" or not.*

# Analog and digital circuit design in 65 nm CMOS: end of the road?

Georges Gielen, Wim Dehaene

Katholieke Universiteit Leuven, ESAT-MICAS
Kasteelpark Arenberg 10
B-3001 Leuven, Belgium

**Abstract**

*This introductory embedded tutorial will give an overview of the design problems at hand when designing integrated electronic systems in nanometer-scale CMOS technologies. First, some general problems that affect circuit design will be addressed such as the increased leakage and variability with scaling technologies. Next, the impact of this on digital circuit design and embedded memories is discussed. Finally, problems bothering embedded analog circuits are presented, such as reducing supply voltages, poor design productivity and signal integrity troubles. Addressing these problems will determine whether the design road ends at CMOS technology marker "65nm" or not.*

## 1. Introduction

With the evolution towards ultra-deep-submicron and nanometer CMOS technologies [1] the design of complex Systems on a Chip (SoC) is emerging in consumer-market applications such as telecom and multimedia. These integrated systems are increasingly mixed-signal designs, embedding high-performance analog or mixed-signal blocks and possibly sensitive RF frontends together with complex digital circuitry (multiple processors, some logic blocks, and several large memory blocks) on the same chip. In addition, the growth of wireless services and other telecom applications increases the need for low-cost highly integrated solutions with very demanding performance specifications.

The use of nanometer CMOS technologies (below 90nm) however also brings along significant challenges for circuit design (both analog and digital). Some of these challenges were never encountered before, while others existed before but have become even stronger limitations today. As an example, in this paper the 65 nm technology node is used but the observations will only be more true if technology scaling is taken to even smaller dimensions.

Technology scaling down to 0.18 μm was based on the following reasoning. Reducing the feature size in the technology front-end (i.e. of the devices) and in the technology back-end (i.e. of the interconnect) combined with the addition of ever more interconnect layers, drastically increased the density of the digital circuits while reducing the intrinsic gate switching delay. The main disadvantage of this is that the power supply voltage must be decreased too, potentially leading to an increase in gate switching delay. However, the threshold voltage ($V_T$) of the MOS transistors is lowered to compensate for this. Assume a full scaling scenario. This means that the scaling factor $1/S$ is the same for all geometry parameters and all voltage parameters. This ideal case leads for standard CMOS gates to a density increase of $S^2$, a decrease of the intrinsic delay of $1/S$, a decrease in power consumption of $1/S^2$ at a constant power density. The noise margin of the logic is decreasing but remains acceptable. This ideal scenario cannot be realized in practice but it shows that for digital circuitry scaling is (was) advantageous: it brings faster, denser logic for less power and an acceptable robustness penalty [2]. For the analog circuits, scaling did not bring much area reduction, but mainly provided the transistors with extra speed, allowing the silicon implementation of RF circuits and high-speed analog blocks like data converters.

However, as technology scaling goes on to 90 nm and below (65nm, 45nm), physical and quantummechanical effects that were previously not relevant become influential or even dominant. An example of this is the advent of no longer negligible leakage currents. Secondly as all geometrical dimensions become smaller, variability of technological parameters also gains in importance. The same absolute tolerance becomes relatively more important when the absolute value of a parameter decreases. This holds for $V_T$, doping levels, widths and lengths, … A $V_T$ variation of 50 mV has more effect on the circuits when $V_T$ is 200 mV compared to a $V_T$ of 700 mV for instance. The central question in this discussion is whether the scaling advantages are still realized under the presence of these new effects. In other words: will the design of circuits in nanometer technologies of 65 nm and beyond still bring the benefits we are used to from the past, or will the design bottlenecks become so large that the design road ends at the 65nm technology node ? This paper will give an introductory overview of problems encountered in nanometer technologies.

The paper is organized as follows. Section 2 gives a more concrete description of the effects that nanometer scaling has on the circuit behavior in general. The 65nm node is used as an example. Section 3 then presents in detail the challenges in digital circuit design, while section 4 focuses on challenges in analog circuit design. Finally, section 5 presents some conclusions.

## 2. General consequences of 65 nm CMOS at the circuit level

Scaling technology to nanometer sizes has brought effects to the surface that were negligible before. The most important ones are discussed here: subthreshold leakage current, gate leakage current, interconnect delay and increased process variability. Their effect on circuit behavior will be described in the next subsections.

### 2.1. Subthreshold leakage current

The current conducted by a MOS transistor when it is off ($V_{GS}$=0) is given by :

$$I_{subthreshold} = I_0 \exp(\frac{-V_T}{nkT/q}) \quad (1)$$

The subthreshold current thus increases exponentially with decreasing $V_T$. Moreover $I_0$ in this equation is inversely proportional to L. Nanometer scaling thus leads to a dramatic increase in leakage current up to the point where it can no longer be ignored in the power consumption nor the functionality of the circuits.

For very short transistors the subthreshold leakage not only increases because L scaling implies $V_T$ scaling. When the drain region of a MOS transistor is close to the source region, the depletion region around the source and drain regions interact with each other. This lowers the potential barrier at the source side. This is usually modeled with an equivalent, $V_{ds}$-dependent, $V_T$ decrease, as illustrated in Fig. 1.

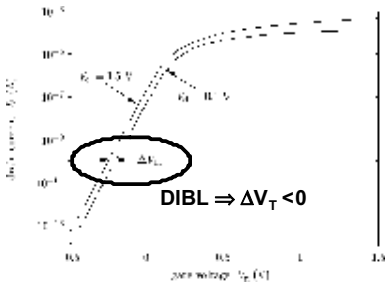

Fig. 1. Subthreshold current dependency on $V_{ds}$ and $V_{gs}$.

### 2.2. Gate leakage current

Before the nanometer era, gate currents were considered as pure dynamic. MOS gates were capacitors and did not conduct DC currents. However, since the advent of oxide thicknesses of only a few nm, current can tunnel through the gate and have a static "leakage" component. The gate leakage current is described by :

$$I_{gate\_leakage} = KW\left(\frac{V_{gb}}{t_{ox}}\right)^2 \exp(-\alpha t_{ox}/V_{gb}) \quad (2)$$

where K and $\alpha$ are fit factors. Remark that gate leakage is only present when there is a voltage across the gate, i.e. when the transistor is on. Subthreshold leakage on the contrary occurs when a transistor is off. Gate leakage currents affect the power consumption and functionality at the circuit level.

Using new technologies not only means scaling of the size. New materials are also introduced. A lot of research attention goes to the introduction of high-k dielectrics for the gate. This would lead to reduced gate leakage because thicker "oxides" can be used without compromising gate capacitance and related to that VT. A similar story holds for the (re-)introduction of metal gates.

### 2.3. Interconnect delay and scaling

A simplified, first-order approximation of interconnect delay is given by :

$$t_{wire} = \frac{rcL^2}{2} = \rho\kappa\left(\frac{L}{\lambda}\right)^2 \quad (3)$$

where r and c are the resistance and capacitance per unit length, $\rho$ and $\kappa$ are the same per unit area, and $\lambda$ is the technology-related wire pitch. This shows that the interconnect delay is inversely proportional to the square of the wire pitch ? of the technology. This pitch scales down with the technology. This means that if the interconnect length (l) and interconnect pitch (?) scale identically, the wire delay will remain constant with technology scaling. As stated in the introduction, the intrinsic gate delay of a technology decreases with technology scaling. This means that the interconnect delay will gain in relative importance when the technologies are scaled down.

Interconnect length and pitch however scale identically for longer interconnects, e.g. busses. The length of these lines has the tendency to remain constant even as technology scales. For that case the interconnect delay becomes dominant even faster.

A similar reasoning can be built up for the power that is consumed due to the interconnect capacitance. Also the relative importance of this contribution to the power consumption is increasing. Also in this case the introduction of new materials could bring some relieve: low-k materials for the isolation between the interconnects for instance, or better conductor materials.

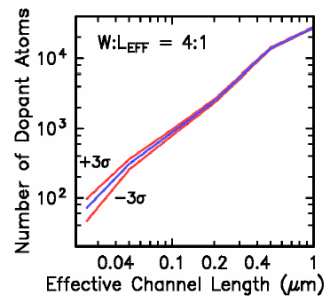

Fig. 2. Number of dopant atoms vs. channel length [3].

### 2.4. Process variability

The tolerance on some technological parameters will increase when the technology scales. There are numerous examples of this, each with a different physical explanation. Two examples will be given here.

The first example is the discrete nature of doping levels. The statistical variation of the number of dopants N varies with $N^{1/2}$. This increases the uncertainty on $V_T$ for small N. Remark that for decreasing channel dimensions, the number of dopants that realizes a given doping level also decreases (see Fig. 2). The random placement of the dopants creating the source and drain also causes an uncertainty on the effective channel length (see Fig. 3). This effect is also enforced as the number of dopants goes down.

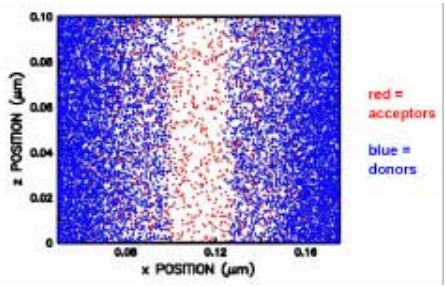

*Fig. 3. Representation of the source/drain dopants in a MOS transistor [4].*

A second example is line edge roughness. As the L of a transistor scales down, the same roughness on the edge of a transistor becomes relatively more important, leading to a larger variation on $L_{eff}$ and thus on the transistor current.

Process variability is split in inter-die and intra-die variations. The latter are also referred to as device matching. Both types of variability are affected by the nanometer scaling. However the methods that can be used at the circuit level to deal with both types are different.

## 3. Digital circuit design challenges in 65 nm

In this section some examples will be given to illustrate the influence of nanometer technology scaling on digital circuit design.

### 3.1. The influence of $V_T$ variability on the energy-delay trade-off

The influence of $V_T$ variations on the delay of a logic gate increases as the technology becomes smaller, as is illustrated in Fig. 4. This is due to the fact that the overdrive voltages ($V_{dd}-V_T$) that are used also become smaller. Therefore the relative importance of ?$V_T$ on $V_{dd}-V_T$ grows.

The effect of this $V_T$ variation is compensated for by designing the circuit such that the delay requirements are still fulfilled in the worst case (largest $V_T$ in this case). This however has a penalty on the energy consumption. The sizing of the circuit, and thus the parasitic capacitance, will be larger than it needs to be when the $V_T$ is not at its largest. However the dynamic energy consumption is determined by $CV_{dd}^2$, independent of $V_T$. The effect of sizing on a circuit's energy consumption is larger than the effect on the circuit's delay. Therefore, the effect of "worst-case oversized" design on the energy consumption of circuits will be significant.

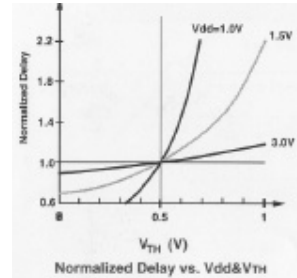

*Fig. 4. Influence of $V_T$ variations on gate delay [5].*

### 3.2. Leakage-aware digital circuits

As explained above, nanometer scaling increases the leakage currents of MOS transistors up to a level where they can no longer be ignored. To deal with this, a range of techniques has been developed and are still under development. A first class of techniques called MTCMOS (multi threshold CMOS) increase the threshold voltage of transistors in the circuit that are not on a critical timing path. This reduces their subthreshold leakage without affecting the speed of the circuit. To apply this technique multiple $V_T$ technologies are required. VTCMOS is usually combined with supply and/or ground switches that switch off the leaky circuits when they are inactive.

A second class of techniques is called VTCMOS (variable threshold CMOS). In this technique the $V_T$ is influenced via the bias of the transistors body. There are a number of strategies to adapt the body bias (see e.g. [6]). There is however one problem with this technique: as technology scales down, the bulk factor becomes smaller. This means that the effect of body bias on $V_T$ reduces, thus limiting the effectiveness of the VTCMOS technique.

The examples given above suggest that there is a point where further scaling of the intrinsic MOS device is not really meaningful anymore. The delay still decreases but the penalty in terms of energy consumption due to leakage and worst case design will become unacceptable. Furthermore, the problems posed to the circuit designer, mainly due to intra-die matching, are hard to deal with. It is more than likely that the circuits needed to solve these problems will also consume power and silicon area…

### 3.3. Increasing interconnect delay and synchronous design

As the relative importance of interconnect delay tends to grow, it becomes harder and harder to maintain the synchronism between different parts of a digital design. In a typical 100 nm technology the max length of a wire is around 2 mm to keep the skew below 20 % of a 1 GHz clock (see Fig. 5). With decreasing interconnect pitches and line widths, this distance will also decrease. This calls for architectural solutions: localization of storage and computation, complemented with a communication strategy that is locally synchronous but globally asynchronous. This will again lead to power and silicon area overhead along with an increased design complexity.

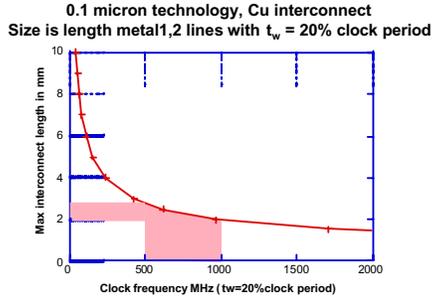

*Fig. 5. Max interconnect length for 20% clock skew as a function of the clock frequency for a typical M1, M2 interconnect in a 100 nm technology.*

## 4. Analog circuit design challenges in 65 nm

In this section some challenges will be discussed that are posed on analog circuit design in nanometer technology scaling.

### 4.1. The influence of reducing supply voltages

For analog circuits technology scaling does not bring large area reductions, because the active area (width times length) of key analog transistors in a circuits is determined by noise or mismatch constraints, both of which limit the dynamic range or accuracy levels that can be achieved. As today's applications tend to increase rather than relax the accuracy requirements, this means that the area of most analog blocks does not really become significantly smaller with technology scaling. In general, both the kT/C thermal noise as the mismatch-induced offset result in a relationship between achievable speed, dynamic range and power consumption of a circuit according to the following formula [7] :

$$\frac{Speed * Accuracy^2}{Power} = techn\, const \quad (4)$$

In the case of thermal noise, the right-hand side constant depends only on temperature; in the case of mismatch it depends on the amount of mismatch in the technology process used. These relationships are plotted for some real technology process in Fig. 6. Clearly, for untrimmed or uncalibrated circuits, the mismatch limit is determining the minimum required power consumption for given speed and dynamic range requirements. The red squares mark real A/D converter designs.

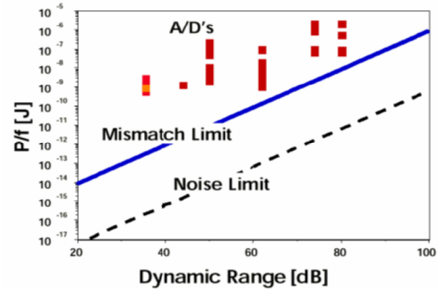

*Fig. 6. Thermal noise and mismatch limit in the power-speed-accuracy trade-off governing analog circuits.*

When the technology scales, then the transistor mismatch improves slightly. Hence, if one wants to exploit the higher speed offered by the scaled technology, then this will come at an increased power consumption for the same level of dynamic range. On the other hand, for fixed speed and fixed accuracy, the power would decrease due to the improved matching. However, this is only true when ignoring the reduction in supply voltage which comes along with employing nanometer technologies. This reduced supply voltage also reduces the input signal range, hence making the constraint on thermal noise and offset even more stringent. As a result, even for fixed speed and fixed accuracy, the power consumption will no longer decrease with scaling, but will remain the same or may even increase slightly (see Fig. 7), according to :

$$\frac{P_1}{P_2} = \frac{1}{m}\frac{t_{ox1}}{t_{ox2}} \quad (5)$$

where $P_1$, $P_2$ are the power consumption in technology 1 and 2 respectively, $t_{ox1}$ and $t_{ox2}$ the oxide thickness in both technologies, and $m$ the ratio of the supply voltages. Hence, there is no real benefit for using nanometer technologies in this case. In addition, the reduced supply voltage also makes it difficult to implement analog circuits, since many circuit techniques like cascading and device stacking become no longer possible.

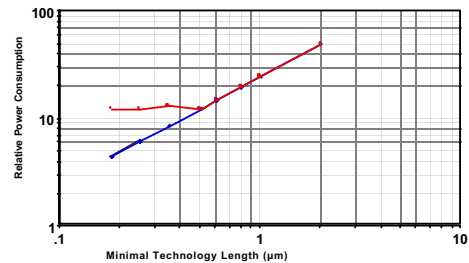

*Fig. 7. Power consumption with technology scaling for analog circuits with fixed speed and accuracy requirements (red curve).*

Also the increased process parameter variability has an impact on analog circuit performance, but analog designers have always had to cope with process tolerances and mismatches, and have been using statistical methods already a long time ago [8].

### 4.2. Boosting analog design productivity

Due to the knowledge-intensive nature of analog design, most analog designs today are still handcrafted manually by analog expert designers, with only a SPICE-like simulation shell and an interactive layout environment (with parameterized procedural device generators) as supporting facilities. This makes the design cycle for analog circuits long and error-prone. Therefore, although analog circuits typically occupy only a small fraction of the total area of mixed-signal ICs and SoCs, their design is often the bottleneck in mixed-signal systems, both in design time and effort as well as test cost, and they are often responsible for design errors and expensive reruns. This handcrafting is also increasingly at odds with the shortening time-to-market constraints of current consumer market products. This explains the growing need observed in industry today for analog CAD tools that boost analog design productivity by assisting designers with fast and first-time-correct design of analog circuits, or even by automating certain tasks or the entire circuit design process where possible.

While the basic level of design abstraction for analog circuits is mainly still the transistor level, commercial CAD tool support for analog cell-level circuit and layout synthesis is emerging. There has been remarkable progress at research level over the past decade, and in recent years several commercial offerings have appeared on the market. [9] offers a fairly complete survey of the domain. Most of the basic techniques in both circuit and layout synthesis today rely on powerful numerical optimization engines coupled to "evaluation engines" that qualify the merit of some evolving analog circuit or layout candidate. These tools have been shown to produce designs comparable or better than manual designs in much shorter design times. As an example, Fig. 8 shows a particle/radiation detector frontend that has been generated with the AMGIE/LAYLA analog synthesis tools [10].

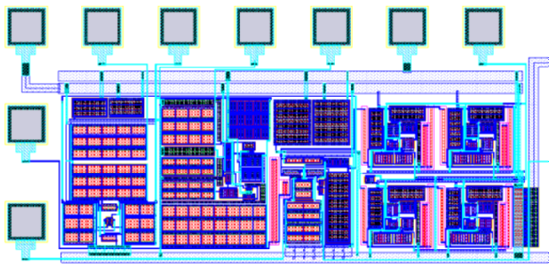

*Fig. 8. Layout of a particle/radiation detector front-end generated with the AMGIE/LAYLA analog synthesis tools [10].*

### 4.3. Mixed-signal signal integrity

A difficult problem in mixed-signal designs, where sensitive analog and RF circuits are integrated on the same die with large digital circuitry, is signal integrity analysis, i.e. the verification of all unwanted signal interactions through crosstalk or couplings at the system level that can cause parametric malfunctioning of the chip. Parasitic signals are generated (e.g. digital switching noise) and coupled into the signal of interest, degrading or even destroying the performance of the analog/RF circuitry. These interactions can come from capacitive or (at higher frequencies) inductive crosstalk, from supply line or substrate couplings, from thermal interactions, from coupling through the package, from electromagnetic interference (EMC/EMI), etc.

Especially the analysis of digital switching noise that propagates through the substrate shared by the analog and digital circuits has received much attention in recent years [11]. At the instants of switching, digital circuitry can inject spiky signals into the substrate, which then will propagate to and be picked up by the sensitive analog/RF circuits. As example, consider a VCO at 2.3 GHz and a digital circuit block (250kgates) running at 13 MHz. As shown on the measurement plot of Fig. 9, the digital clock is visible as FM modulation around the VCO frequency and may cause conflicts with out-of-band emission requirements [12].

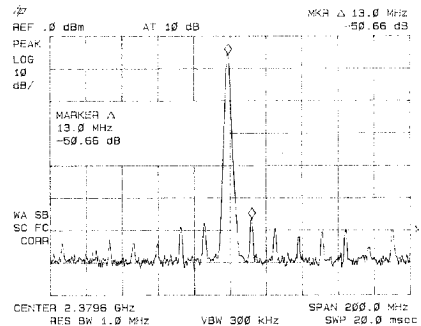

*Fig. 9. Measured FM modulation due to substrate switching noise coupling [12].*

In recent years research has been going on to find efficient yet accurate techniques to analyze these problems. For the noise propagation through the substrate, typically finite difference methods or boundary element methods are used to solve for the substrate potential distribution due to injected noise sources [11], allowing to simulate the propagation of digital switching noise injected in the substrate to sensitive analog nodes elsewhere in the same substrate. This propagation analysis then has to be combined with an analysis of the (signal-dependent) digital switching activity to know the actual (time-varying) injected signals, and with an analysis of the impact of the local substrate voltage variations on the analog/RF circuit performance (e.g. the reduction of the effective number of bits of an embedded analog-to-digital converter) in order to cover the entire problem.

As an example, the SWAN methodology determines the switching noise that is generated by the digital circuits in a system, by a-priori characterizing every cell in a digital standard cell library with a macromodel that includes the current injected in the substrate due to an input transition, and then calculating the total injection of a complex system by combining the contributions of all switching cells over time depending on the event information obtained from a VHDL simulation of the system [13]. Fig. 10 for example shows the comparison between time-domain SWAN simulations and measurements on a large experimental WLAN SoC with 220k gates, that contains a scalable OFDM-WLAN baseband modem and a low-IF digital IQ (de)modulator, fabricated in a 3.3 V 0.35 µm CMOS 2P5M process on an EPI-type substrate. Compared to the measurements, the simulated substrate-noise voltage from zero to 100 ns is within an error of 20% in its RMS value and within an error of 4% in its peak-to-peak value, which is a very good result for a difficult crosstalk effect like substrate noise couplings. Techniques to analyse the impact of this on the performance of the embedded analog blocks are also being developed, but still require further work [11].

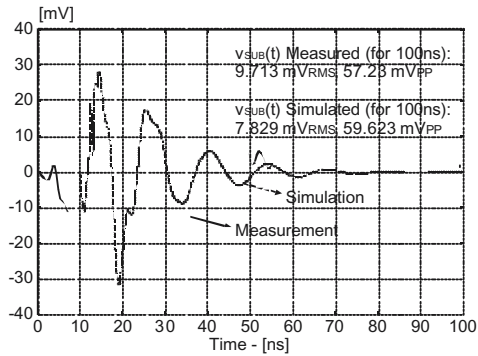

*Fig. 10. Measured and SWAN [13] simulated substrate noise in an experimental 220k-gates WLAN SoC.*

In future nanometer technologies, however, also other signal integrity problems will show up that need to be analysed and modeled. These include electromagnetic interactions, EMC/EMI, etc.

## 5. Conclusions

This paper has presented an overview of the challenges at hand when designing integrated electronic systems in nanometer-scale CMOS technologies. Key problems that affect circuit design with scaling technologies include increased leakage and variability. Next, the impact of this on digital circuit design has been discussed. Finally, problems bothering embedded analog circuits have been presented, such as reducing supply voltages, poor design productivity and signal integrity troubles. Whether we manage to address these problems or not will determine whether the design road ends at CMOS technology marker "65nm" or not.

## 6. References

[1] "International Technology Roadmap for Semiconductors 2003," http://public.itrs.net.
[2] J. Rabaey, B. Nikolic, A. Chandrakasan, "Digital integrated circuits ($2^{nd}$ edition)", Prentice Hall, 2002.
[3] D. Frank,"Parameter variations in sub-100nm technology," ISSCC microprocessor workshop, San Fransico, February 2004.
[4] D. Frank, P. Wong, "Simulation of Stochastic Doping Effects in Si MOSFETs," IWCE, 2000.
[5] T. Kobayashi *et al.*, "Self-adjusting threshold-voltage scheme (SATS) for low-voltage high-speed operation,"proc. CICC, pp.271-274, May 1994.
[6] T. Sakurai, "Adaptive Circuit Techniques for Managing Variations," ISSCC microprocessor workshop, San Fransico, February 2004.
[7] P. Kinget, M. Steyaert, "Impact of transistor mismatch on the speed-accuracy-power trade-off of analog CMOS circuits," proc. CICC, pp. 333-336, May 1996.
[8] S. Director, P. Feldmann, K. Krishna, "Statistical integrated circuit design," IEEE Journal of Solid-State Circuits, Vol. 28, No. 3, pp. 193-202, March 1993.
[9] G. Gielen, R. Rutenbar, "Computer-aided design of analog and mixed-signal integrated circuits," Proceedings of the IEEE, Vol. 88, No. 12, pp. 1825-1854, December 2000.
[10] G. Van der Plas, et al., "AMGIE - A synthesis environment for CMOS analog integrated circuits," IEEE Transactions on Computer-Aided Design of Integrated Circuits and Systems, Vol. 20, No. 9, pp. 1037-1058, September 2001.
[11] S. Donnay, G. Gielen (editors), "Analysis and reduction techniques for substrate noise coupling in mixed-signal integrated circuits," European Mixed-Signal Initiative for Electronic System Design, Kluwer Academic Publishers, 2003.
[12] D. Leenaerts, G. Gielen, R. Rutenbar, "CAD solutions and outstanding challenges for mixed-signal and RF IC design," proceedings ICCAD, pp. 270-277, November 2001.
[13] M. van Heijningen, M. Badaroglu, S. Donnay, G. Gielen, H. De Man, "Substrate noise generation in complex digital systems: efficient modeling and simulaton methodology and experimental verification," IEEE Journal of Solid-State Circuits, Vol. 37, No. 8, pp. 1065-1072, August 2002.